\documentclass[twocolumn,pra,superscriptaddress,showpacs]{revtex4}
\usepackage[utf8]{inputenc}
\usepackage{amsmath}
\usepackage{amssymb}
\usepackage{graphicx}
\usepackage{esint}
\usepackage[unicode=true,pdfusetitle,
 bookmarks=true,bookmarksnumbered=false,bookmarksopen=false,
 breaklinks=false,pdfborder={0 0 1},backref=section,colorlinks=false]
 {hyperref}
\hypersetup{
 colorlinks,citecolor=blue}
\usepackage{breakurl}

\makeatletter
\@ifundefined{textcolor}{}
{%
 \definecolor{BLACK}{gray}{0}
 \definecolor{WHITE}{gray}{1}
 \definecolor{RED}{rgb}{1,0,0}
 \definecolor{GREEN}{rgb}{0,1,0}
 \definecolor{BLUE}{rgb}{0,0,1}
 \definecolor{CYAN}{cmyk}{1,0,0,0}
 \definecolor{MAGENTA}{cmyk}{0,1,0,0}
 \definecolor{YELLOW}{cmyk}{0,0,1,0}
}

\usepackage{subfigure}\usepackage{epsfig}\usepackage{amsfonts}\usepackage{mathrsfs}\usepackage{CJK}\usepackage[toc,page,title,titletoc,header]{appendix}

\makeatother

\begin{document}

\title{Orbital Feshbach Resonance with Small Energy Gap between Open and
Closed Channels}

\author{Yanting Cheng}

\affiliation{Institute for Advanced Study, Tsinghua University, Beijing, 100084,
China}

\author{Ren Zhang}

\affiliation{Institute for Advanced Study, Tsinghua University, Beijing, 100084,
China}

\author{Peng Zhang}

\affiliation{Department of Physics, Renmin University of China, Beijing, 100872,
China}

\affiliation{Beijing Computational Science Research Center, Beijing, 100084, China}

\affiliation{Beijing Key Laboratory of Opto-electronic Functional Materials \&
Micro-nano Devices (Renmin University of China)}

\begin{abstract}
Recently a new type of Feshbach resonance, i.e., orbital Feshbach
resonance (OFR) was proposed for the ultracold alkali-earth (like)
atoms, and experimentally observed in the ultracold gases of $^{{\rm 173}}$Yb
atoms. Unlike most of the magnetic Feshbach resonances of ultracold
alkali atoms, when the OFR of $^{{\rm 173}}$Yb atoms appears, the energy
gap between the thresholds of the open channel (OC) and the closed
channel (CC) is much smaller than the characteristic energy of the
inter-atomic interaction, i.e., the van der Waals energy. In this
paper we study the OFR in the systems with small CC-OC threshold
gap. We show that in these systems the OFR can be induced by 
the coupling between the OC and either an isolated bound state of the
CC or the scattering states of
the CC.
Moreover, we also show that in each case the two-channel
Huang-Yang pesudopoential is always applicable
for the approximate calculation of the low-energy scattering amplitude.
Our results imply that in the theoretical calculations for these systems it is appropriate to take into account
the contributions from the scattering states of the CC.

\end{abstract}


\maketitle

\section{introduction}

Feshbach resonance \cite{fr} exists in many kinds of ultracold gases,
and can be used as a power tool for controlling the interaction between
ultracold atoms \cite{chinrmp}. For instance, in the ultracold gases
of alkali atoms, when the relative motional state of two atoms in
 the open channel (OC) is near resonant to
a bound state of the closed channel (CC) with higher threshold
energy, magnetic Feshbach resonance (MFR) can
be induced by the hyperfine coupling between different channels \cite{paulrmp}.
With the help of this resonance effect, one can precisely control
the scattering length between these two atoms by magnetically changing
the inter-channel energy difference. In almost all the current experiments
of ultracold alkali atoms, when the MFR appears, the gap between the
threshold energies of the OC and the CC is as high as $10^{8}$Hz$-10^{9}$Hz \cite{chinrmp}. As a result
of this large energy gap, the resonance is usually induced by the coupling
between the OC and a single bound state (or several bound
states with similar energies) of the CC. The contribution
of the scattering states of the CC for the resonance can
be neglected.

For ultracold alkali-earth (like) atoms, recently we
proposed a new type of Feshbach resonance, i.e., orbital Feshbach
resonance (OFR) \cite{ourprl}. This resonance can occur in the scattering
between two alkali-earth (like) atoms in different electronic orbital
and nuclear spin states. The OFR is a result of the spin-exchange
interaction \cite{spinexchange1,spinexchange2,spinexchange3} and
the Zeeman effect \cite{zeeman} in such system, and can be used for
magnetically controlling the interaction between alkali-earth (like)
atoms. The OFR has been experimentally observed in ultracold $^{173}{\rm Yb}$
atoms which are in the $^{1}{\rm S}_{0}$ and $^{3}{\rm P}_{0}$ electronic
orbital states with different quantum numbers of  nuclear spins \cite{ofrexp1,ofrexp2}.
In these experiments, when the OFR occurs, the energy gap between
the thresholds of the CC and the OC is about $2\times10^{5}$Hz \cite{ofrexp1,ofrexp2}.
This energy gap is not only much smaller than the CC-OC threshold gap of the
MFR of ultracold alkali atoms, but also much smaller than the characteristic
energy (i.e., the van der Waals energy) of the interaction potential
between these $^{173}{\rm Yb}$ atoms, which is about $1.8\times10^{7}$Hz
\cite{spinexchange3}.

This paper is to address the effect of this small energy gap in OFR. Our results can be summarized as follows:
\begin{itemize}
\item[(i)] For the systems with small CC-OC threshold gap,
 the effects contributed by the scattering states 
 of the CC may be very important. As a result, the OFR
can be induced by the coupling of either of the following two types:
\begin{itemize}
\item[(A)] the coupling between the OC
and an isolated bound state of the CC.
\item[(B)] the coupling between
the OC and the scattering states of the CC \cite{frwobs1,frwobs2}. 
\end{itemize}
\item[(ii)]  In each of the above cases, the two-channel Huang-Yang pseudopotential
\cite{ourprl,huangyang} (i.e., the pseudopotential used in our previous
work \cite{ourprl}) is always applicable
for the approximate calculation of the low-energy scattering amplitude.
\end{itemize}
Our results imply that in the few-body or many-body calculations
for the systems with small CC-OC threshold gap,
it is appropriate to take into account the contributions from
the scattering states of the CC.


This paper is organized as follows. 
In Sec. \ref{sec:hamilton}, we derive the Hamiltonian for alkali-earth (like) atoms with different orbital and spin, and introduce the OFR.
In Sec. III we illustrate result (i) with a square-well model by analyzing three different cases. 
In Sec. \ref{HY} we investigate the applicability of the two-channel Huang-Yang pseudopotential and derive the result (ii). Some discussions are given in Sec. \ref{sec:dis}. In the appendix
we provide a brief explanation for the principle of the Feshbach
resonance induced by the coupling between the OC and the scattering
states of the CC \cite{frwobs1,frwobs2} .

\begin{figure}
\label{fig:energy-level}
\includegraphics[width=7.5cm]{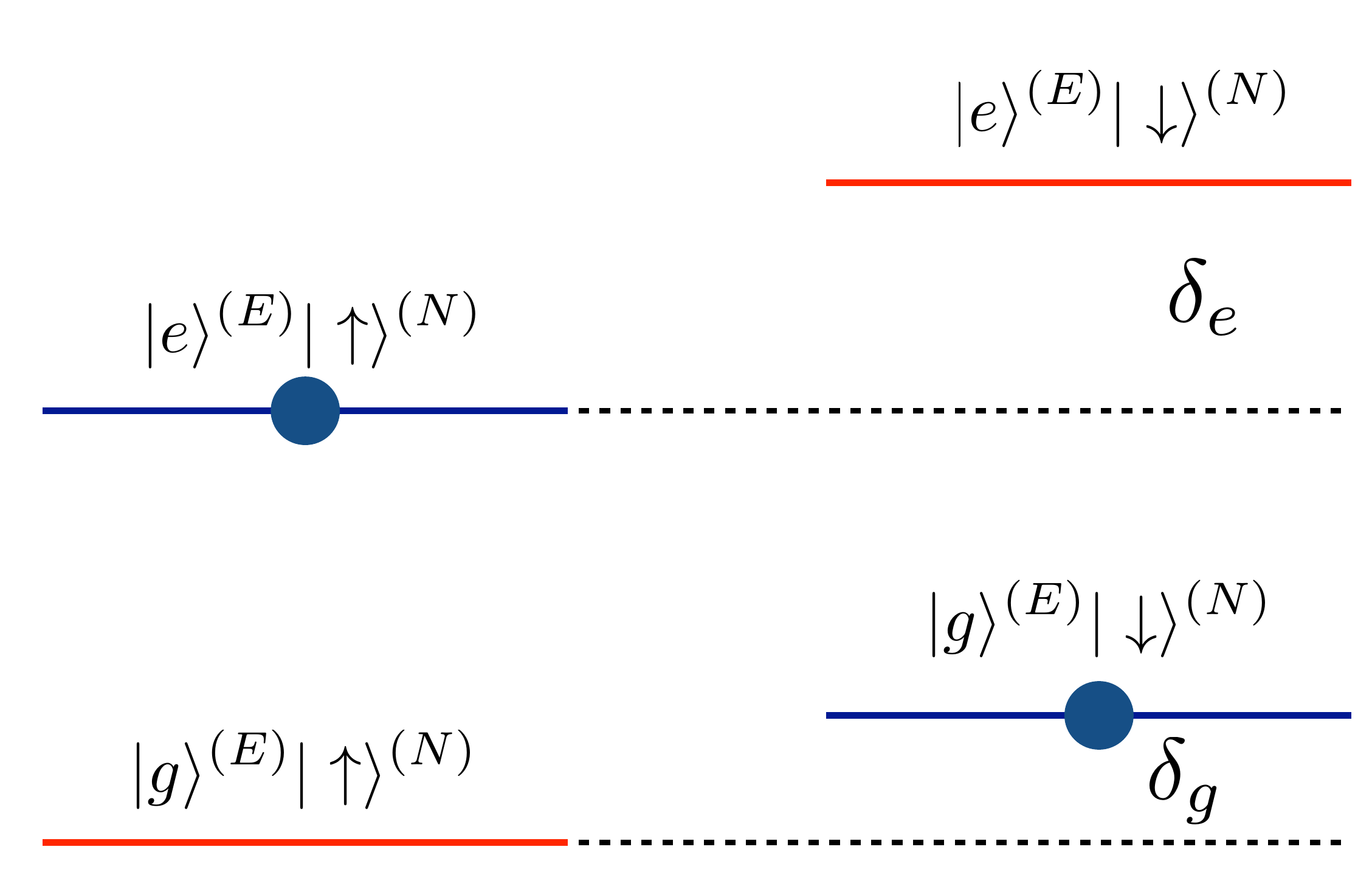} \caption{(color online) Energy level diagram of a single atom. Here $\delta_{e}$
is the Zeeman-energy difference between the states $|e\rangle^{({\rm E})}|\downarrow\rangle^{({\rm N})}$
and $|e\rangle^{({\rm E})}|\uparrow\rangle^{({\rm N})}$, and $\delta_{g}$
is the one between states $|g\rangle^{({\rm E})}|\downarrow\rangle^{({\rm N})}$
and $|g\rangle^{({\rm E})}|\uparrow\rangle^{({\rm N})}$. The Zeeman
energy difference $\delta$ in Eq. (\ref{h}) can be expressed as
$\delta=\delta_{e}-\delta_{g}$. In our system with $\delta>0$, the
open channel is the state $|o\rangle$ with one atom in $|e\rangle^{({\rm E})}|\uparrow\rangle^{({\rm N})}$
and one atom in $|g\rangle^{({\rm E})}|\downarrow\rangle^{({\rm N})}$.}
\end{figure}

\section{OFR of alkali-earth (like) atoms}
\label{sec:hamilton}

We consider the scattering of two fermionic alkali-earth (like) atoms
in the $^{1}{\rm S}_{0}$ and $^{3}{\rm P}_{0}$ electronic orbital
states with different quantum numbers of nuclear spin. We denote
the electronic states $^{1}{\rm S}_{0}$ and $^{3}{\rm P}_{0}$ for
the $i$th atom ($i=1,2$) as $|g\rangle_{i}^{({\rm E})}$ and $|e\rangle_{i}^{({\rm E})}$,
respectively, and denote the nuclear spin states for the $i$-th
atom as $|\uparrow\rangle_{i}^{({\rm N})}$ and $|\downarrow\rangle_{i}^{({\rm N})}$
(Fig. 1). Here the superscript E and N denote the electronic orbital and nuclear spin degree of freedom, respectively.  We further define $|o\rangle$ as the state where one atom
is in $|g\rangle^{({\rm E})}|\downarrow\rangle^{({\rm N})}$ and the
other one is in $|e\rangle^{({\rm E})}|\uparrow\rangle^{({\rm N})}$,
and $|c\rangle$ as the state with one atom being in $|g\rangle^{({\rm E})}|\uparrow\rangle^{({\rm N})}$
and the other one being in $|e\rangle^{({\rm E})}|\downarrow\rangle^{({\rm N})}$,
i.e., 
\begin{widetext}
\begin{eqnarray}
|o\rangle & \equiv & |g,\downarrow;e,\uparrow\rangle\equiv\frac{1}{\sqrt{2}}\left[|g\rangle_{1}^{({\rm E})}|\downarrow\rangle_{1}^{({\rm N})}|e\rangle_{2}^{({\rm E})}|\uparrow\rangle_{2}^{({\rm N})}-|e\rangle_{1}^{({\rm E})}|\uparrow\rangle_{1}^{({\rm N})}|g\rangle_{2}^{({\rm E})}|\downarrow\rangle_{2}^{({\rm N})}\right],\label{os}\\
|c\rangle & \equiv & |g,\uparrow;e,\downarrow\rangle\equiv\frac{1}{\sqrt{2}}\left[|g\rangle_{1}^{({\rm E})}|\uparrow\rangle_{1}^{({\rm N})}|e\rangle_{2}^{({\rm E})}|\downarrow\rangle_{2}^{({\rm N})}-|e\rangle_{1}^{({\rm E})}|\downarrow\rangle_{1}^{({\rm N})}|g\rangle_{2}^{({\rm E})}|\uparrow\rangle_{2}^{({\rm N})}\right].\label{cs}
\end{eqnarray}
\end{widetext}
The Hamiltonian of such system can be written as ($\hbar=m=1$, with $m$ being the
single-atom mass) 
\begin{equation}
\hat{H}=-\nabla_{{\bf r}}^{2}+\delta|c\rangle\langle c|+\hat{U}\equiv\hat{H}_{0}+\hat{U},\label{h}
\end{equation}
where $\hat{H}_{0}$ and $\hat{U}$ are the free Hamiltonian for the
two-atom relative motion and the inter-atomic interaction, respectively.
Here ${\bf r}$ is the relative position of these two atoms,
and $\delta$ is the Zeeman energy difference of the states $|c\rangle$
and $|o\rangle$. It can be expressed as $\delta=(\Delta g)\mu_{{\rm B}}B(m_{\downarrow}-m_{\uparrow})$,
with $B$ being the magnetic field, $\mu_{{\rm B}}$ being the Bohr's magneton,
$\Delta g$ being the difference of the Landé g-factors corresponding
to states $|e\rangle^{({\rm E})}$ and $|g\rangle^{({\rm E})}$, and $m_{\uparrow}$ ($m_{\downarrow}$) being
the quantum number of nuclear spin for states $|\uparrow\rangle^{({\rm N})}$ ($|\downarrow\rangle^{({\rm N})}$).
Without loss of generality, we assume $\Delta g>0$ and thus $\delta>0$.
In such system the inter-atomic interaction $\hat{U}$ can be expressed as \cite{spinexchange1,spinexchange2,spinexchange3}
\begin{equation}
\hat{U}=U^{(+)}(r)|+\rangle\langle+|+U^{(-)}(r)|-\rangle\langle-|,\label{u}
\end{equation}
where the states $|+\rangle$ and  $|-\rangle$ are defined as
\begin{eqnarray}
|\pm\rangle&\equiv&\frac{1}{2}\left[|g\rangle_{1}^{({\rm E})}|e\rangle_{2}^{({\rm E})}\pm|e\rangle_{1}^{({\rm E})}|g\rangle_{2}^{({\rm E})}\right]\nonumber \nonumber\\
&&\otimes\left[|\uparrow\rangle_{1}^{({\rm N})}|\downarrow\rangle_{2}^{({\rm N})}\mp|\downarrow\rangle_{1}^{({\rm N})}|\uparrow\rangle_{2}^{({\rm N})}\right]\nonumber\\
&=&\frac{1}{\sqrt{2}}\left(|c\rangle\mp|o\rangle\right),\label{pm}
\end{eqnarray}
respectively, and 
$U^{(\pm)}(r)$ is the potential curve with respect to state
$|\pm\rangle$. 

\begin{figure}
\label{fig:vw}
\includegraphics[width=8.5cm]{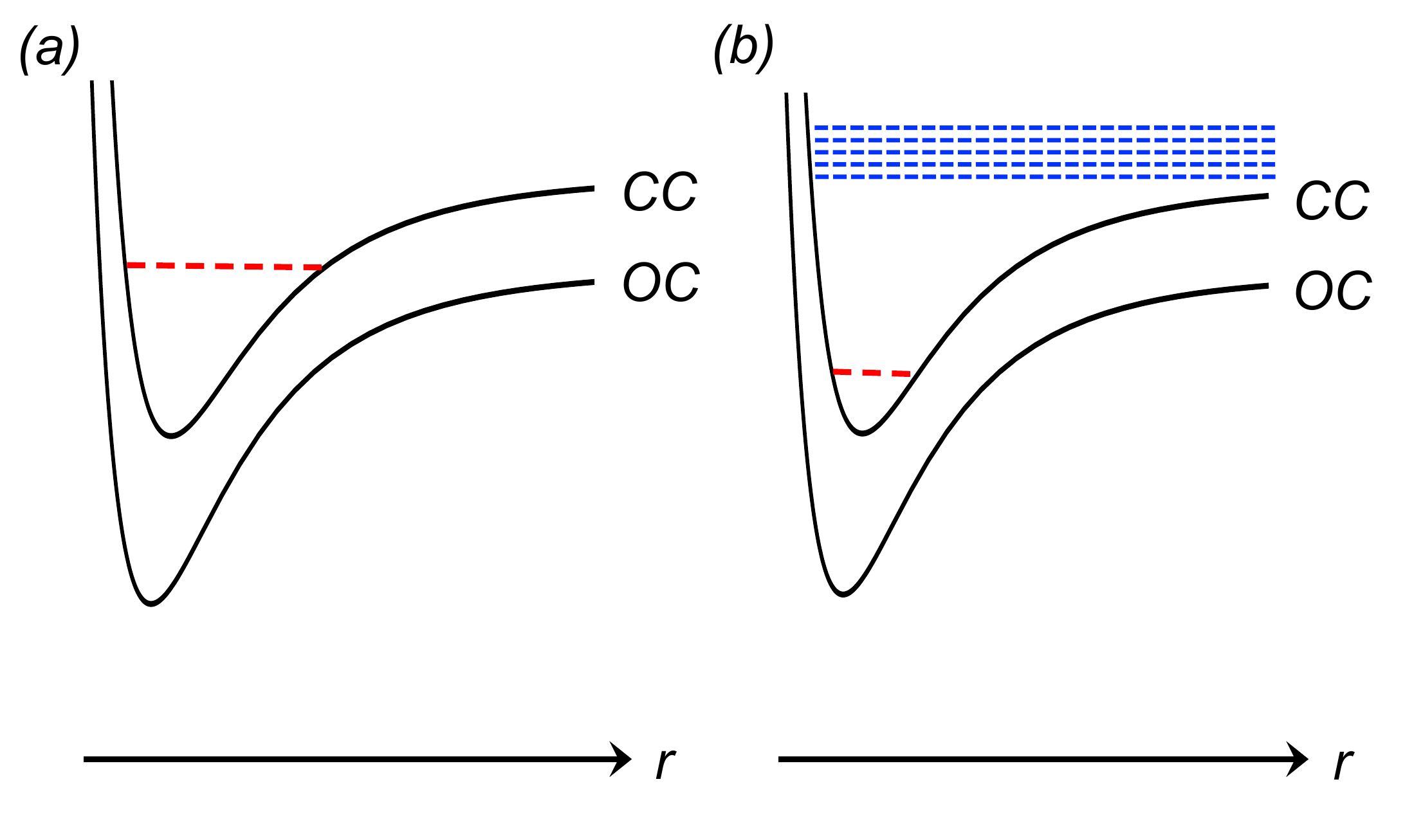} \caption{(color online) Schematic diagram for the Feshbach resonances. Here
$r$ is the inter-atomic distance and the black solid lines are the
potential curves of the OC and the CC. (a) The Feshbach resonance
induced by the coupling between the OC and an isolated bound state
(red dashed line) of the CC. (b) The Feshbach resonance induced
by the coupling between the OC and the scattering states (blue dashed
dotted lines) of the CC. In this case all the bound states (red dashed
lines) of the CC are far-off resonant to the threshold of the OC.}
\end{figure}

It is clear that the  free Hamiltonian $\hat{H}_0$, which governs the physics in the long-distance region where the two atoms are far away from each other, is diagonal in the basis $|o\rangle$ and $|c\rangle$. On the other hand, the interaction potential $\hat{U}$, which is very important when the inter-atom distance is short, is diagonal in another bases $|+\rangle$ and $|-\rangle$. In the conventional treatment \cite{chinrmp}, we always take the same bases in the short-distance region as in the long-distance region, so that the kinetic energy takes the same form in each region. Therefore, we define the OC and CC as the scattering channels corresponding to $|o\rangle$ and $|c\rangle$, respectively. In this bases, the interaction potential is non-diagonal, and can be re-expressed as
\begin{equation}
\hat{U}=\sum_{i,j=o,c}U_{ij}(r)|i\rangle\langle j|,\label{u01}
\end{equation}
where
\begin{equation}
U_{oo}(r)=U_{cc}(r)=\frac{1}{2}\left[U^{(+)}(r)+U^{(-)}(r)\right]\label{uoo}
\end{equation}
and 
\begin{equation}
U_{oc}(r)=U_{co}(r)=\frac{1}{2}\left[U^{(-)}(r)-U^{(+)}(r)\right]\label{uoc}
\end{equation}
Here  $U_{oo}(r)$ and $U_{cc}(r)$ can be viewed as the intra-channel
potential for the OC and the CC, respectively, while $U_{oc}(r)$
and $U_{co}(r)$ can be viewed as the inter-channel coupling between the OC and the
CC. 

We study the scattering length $a_{{\rm s}}$ between two atoms in the OC $|o\rangle$.
This scattering length is defined via the relation
\begin{eqnarray}
|\Psi(r\rightarrow\infty)\rangle\propto\left(\frac{1}{r}-\frac{1}{a_{\rm s}}\right)|c\rangle.
\end{eqnarray}
Here $|\Psi({\bf r})\rangle$ is the scattering wave function of the threshold scattering 
of two atoms in in the OC, i.e., the wave function which satisfies $H|\Psi({\bf r})\rangle=0$.
Due to the coupling $U_{oc}(r)$ between the OC and the CC,
 $a_{\rm s}$ is a function of gap $\delta$ between the thresholds of these two channels.
 In some systems,  $a_{\rm s}$ diverges when this threshold gap
 is tuned to some particular value. That is the OFR.

In this paper we consider the systems where the CC-OC threshold gap $\delta$
is much smaller than the characteristic energy $E_{\ast}$ of the
interaction potential, i.e., the systems with
\begin{eqnarray}
\delta\ll E_{\ast}. \label{con1}
\end{eqnarray}
Here the characteristic energy $E_{\ast}$ are defined as
\begin{eqnarray}
E_{\ast}=\frac{1}{r_{\ast}^2},\label{east}
\end{eqnarray}
where $r_{\ast}$ is the characteristic length
of the interaction potential $U^{(\pm)}(r)$ and satisfies $U^{(\pm)}(r\gtrsim r_{\ast})\approx 0$.
For a realistic ultracold gas of alkali-earth (like) atoms, $r_{\ast}$ can be chosen as the van der Waals radius $R_{{\rm vdW}}$
which is related to the asymptotic  behavior of interaction by
\begin{equation}
U^{(\pm)}(r\rightarrow\infty)=-\frac{(2R_{{\rm vdW}})^{4}}{r^{6}}.\label{abar}
\end{equation}


\section{OFR induced by the couplings of types (A) and (B)}

In this section, using a simple square-well model \cite{swpaper1,swpaper2,swpaper3} 
we illustrate that the OFR in the systems under the condition (\ref{con1}) can be induced by the coupling between the OC and
either the scattering states of the CC (Fig. 2a) \cite{frwobs1,frwobs2} or an isolated bound state of the CC (Fig. 2b), i.e.,
the couplings of either type (A) or type (B) we introduced in Sec. I.

In our model $U^{(+)}(r)$ and $U^{(-)}(r)$ in Eq.~(\ref{u}) are
square-well potentials which satisfy (Fig. 3)
\begin{equation}
U^{(\pm)}(r)=\left\{ \begin{array}{c}
-u^{(\pm)},\ r<b\\
0,\ r>b
\end{array}\right..\label{usw}
\end{equation}
where $b$ is the range of the square-well potential.
For simplicity, here we only consider the potentials $U^{(\pm)}(r)$
with at most two bound states. In this model we have $r_{\ast}=b$ and thus
 $E_{\ast}=b^{-2}.$

\begin{figure}
\label{fig:sw-model}
\includegraphics[width=8cm]{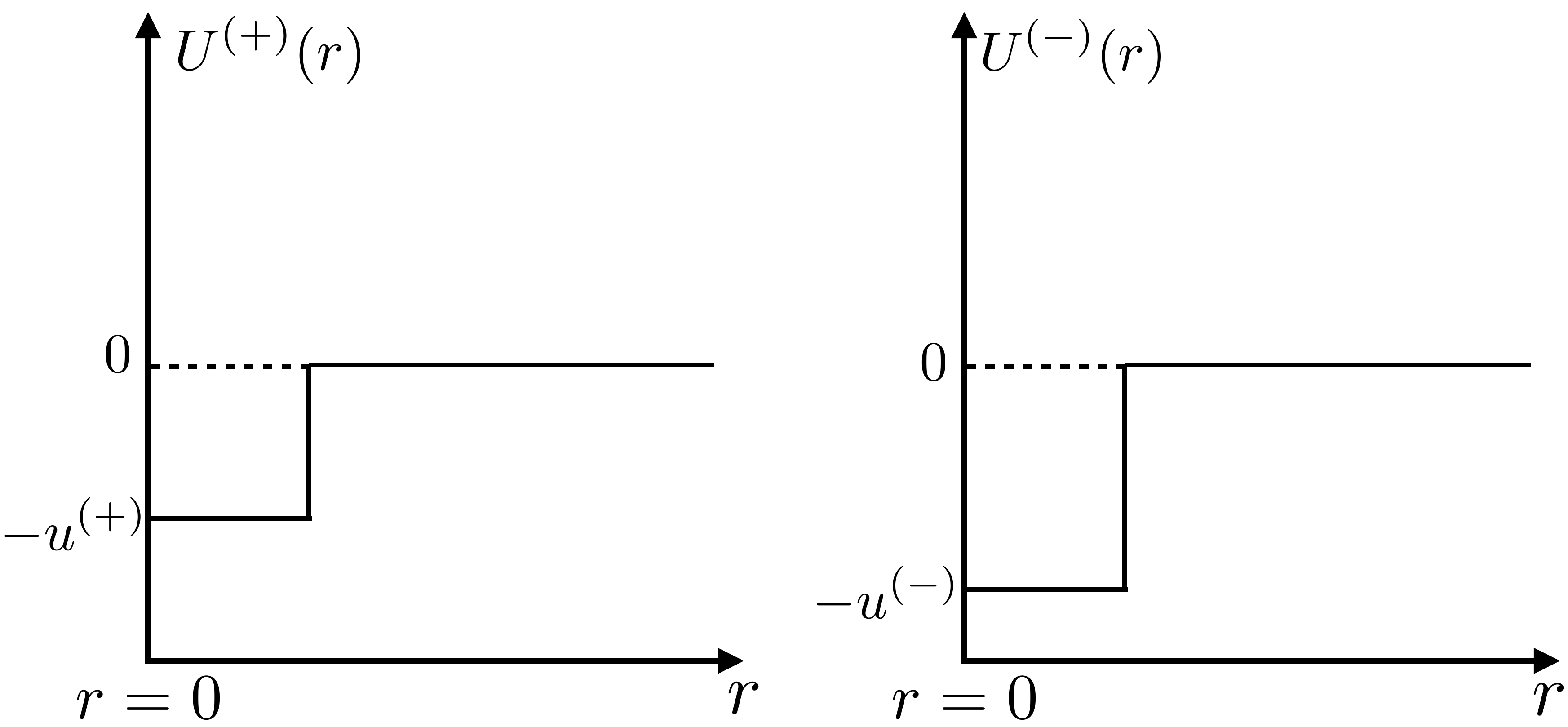} \caption{(color online) The square-well model for the potentials $U^{(+)}(r)$
and $U^{(-)}(r)$.}
\end{figure}

\begin{figure}
\begin{minipage}[b]{0.5\textwidth}
\includegraphics[width=9cm]{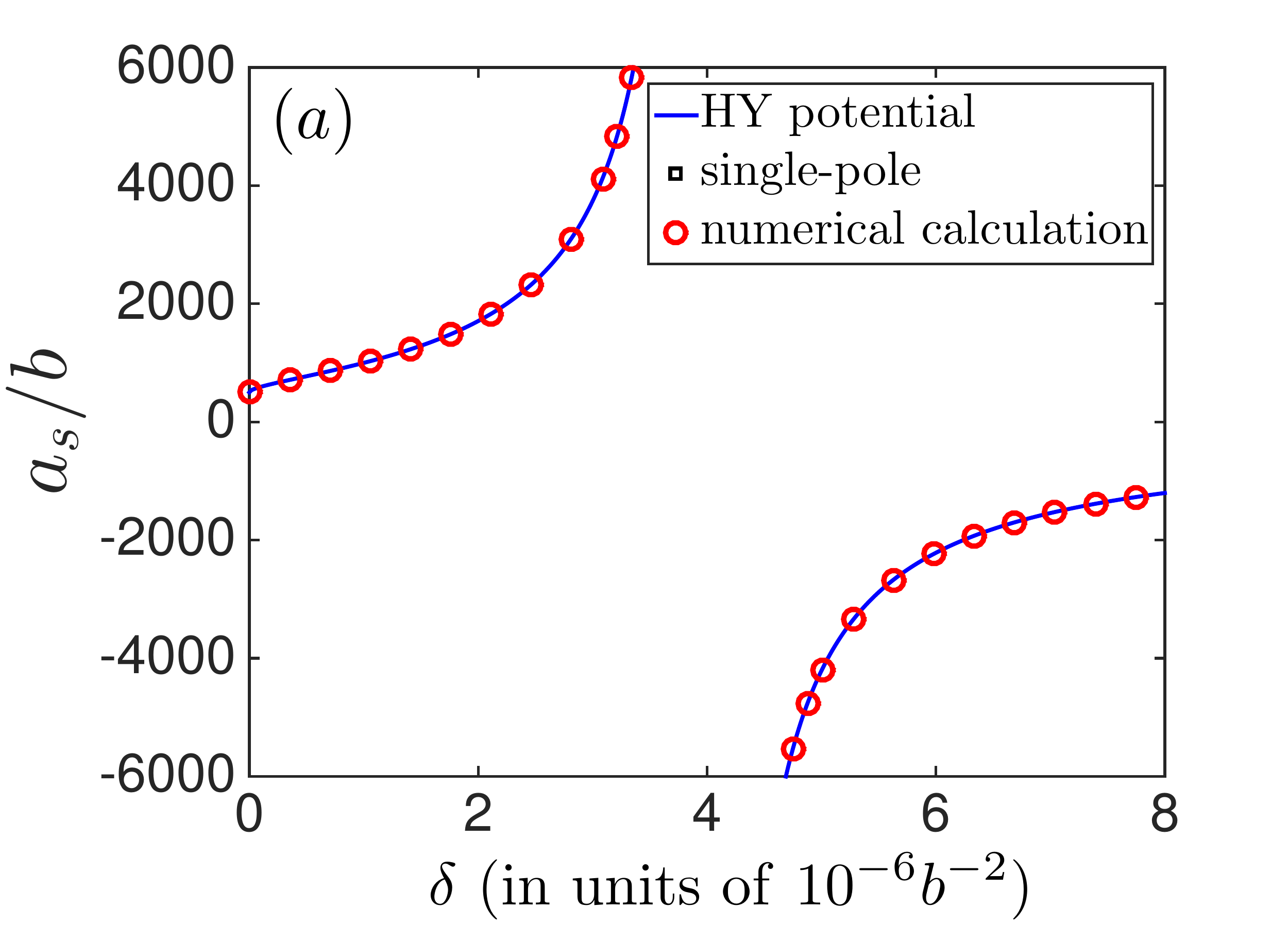} \\
\includegraphics[width=9cm]{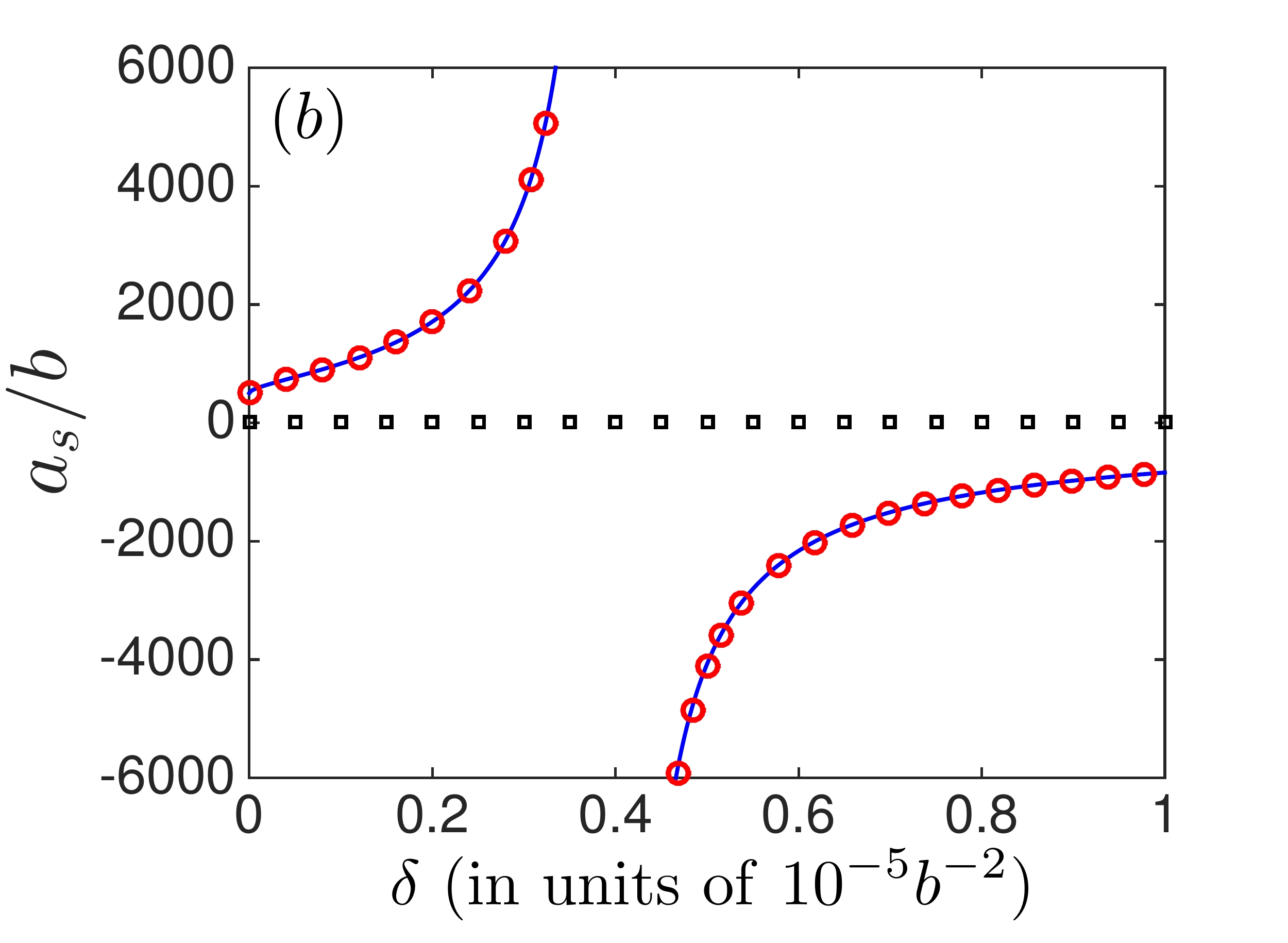}\\
\includegraphics[width=9cm]{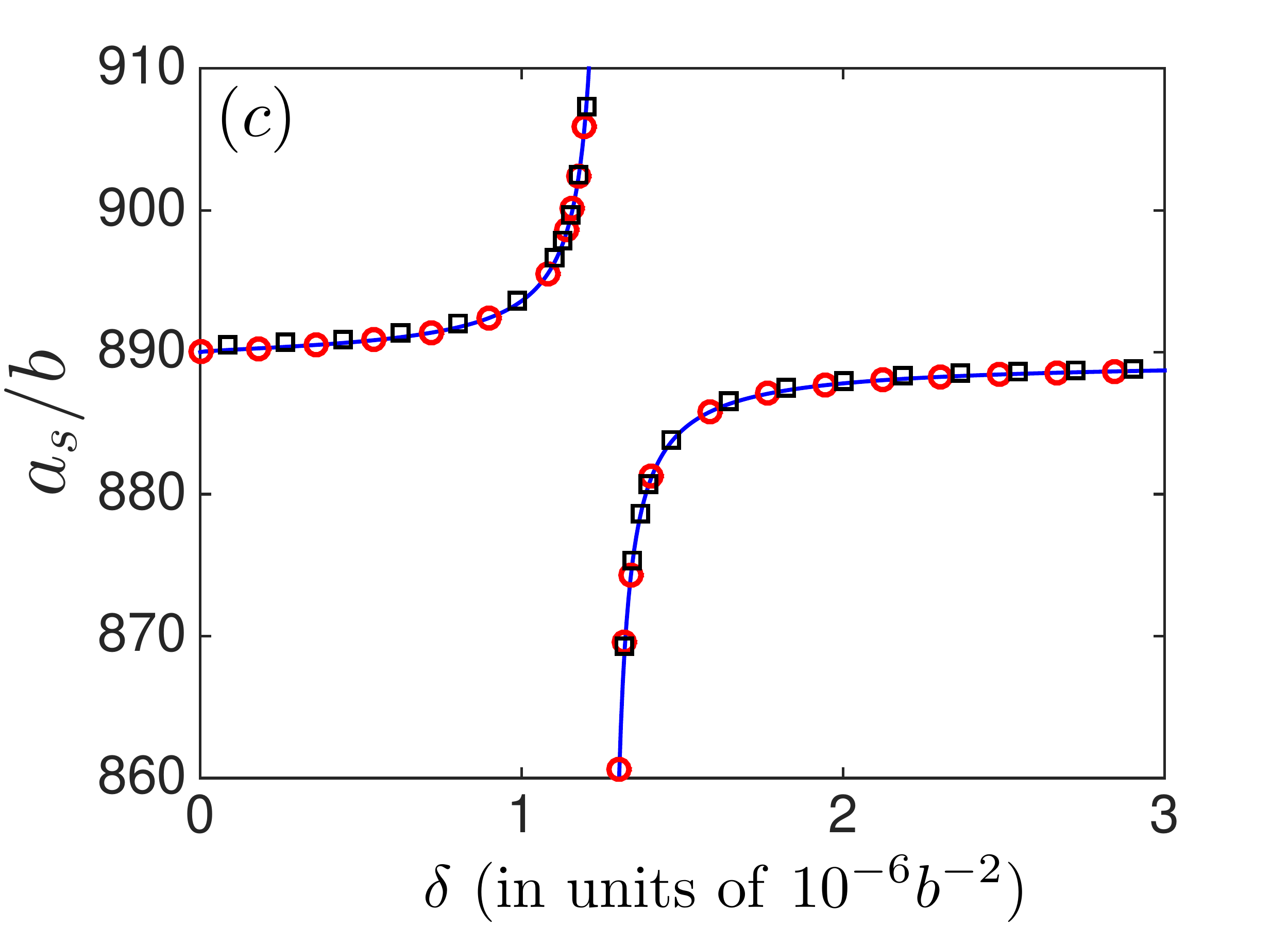}
\end{minipage}
 \caption{(color online) The scattering length $a_{{\rm s}}$ of the square-well model. We show the results given by exact numerical calculation
(red circle) and the two-channel Huang-Yang pseudopotential (blue
solid line). In (a) and (c) we also show the single-pole approximation
(black square). Here we consider the cases with (a):  $a^{(+)}=1000b$ ($u^{(+)}\approx2.47b^{-2}$) and $a^{(-)}=0.5b$
($u^{(-)}\approx-3.67b^{-2}$); 
(b):$a^{(+)}=0.15b$
($u^{(+)}\approx19.8b^{-2}$) and $a^{(-)}=1000b$ ($u^{(+)}\approx22.2b^{-2}$);(c): $a^{(+)}=870b$ ($u^{(+)}\approx2.4697b^{-2}$)
and $a^{(-)}=900b$ ($u^{(-)}\approx2.46963b^{-2}$).}
\end{figure}

In this section we calculate the scattering lengt $a_{{\rm s}}$ between two atoms in the OC with both the exact
numerical calculation for the above square-well potential and the single-pole
approximation. In the  single-pole
approximation only the contribution from a
single bound state of the CC, which is nearest resonant to the threshold
of the OC, is taken into account. The contributions from all the other
bound states as well as the scattering states of the CC are neglected in this approximation.
Under this approximation, the scattering length $a_{{\rm s}}$ is
given by \cite{paulrmp} 
\begin{equation}
a_{{\rm s}}=a_{{\rm bg}}+\frac{2\pi^{2}|w|^{2}}{|\epsilon_{b}|-\delta-\epsilon_{0}},\label{asp}
\end{equation}
with 
\begin{equation}
\epsilon_{0}=\int d{\bf r}d{\bf r}'\phi_{b}({\bf r})^{\ast}U_{co}(r)G_{{\rm bg}}({\bf r},{\bf r}')U_{oc}(r')\phi_{b}({\bf r}')\label{e0-1}
\end{equation}
and 
\begin{equation}
w=\int d{\bf r}\phi_{b}({\bf r})^{\ast}U_{co}(r)\psi_{{\rm bg}}({\bf r}).\label{w-1}
\end{equation}
Here $\phi_{b}({\bf r})$ and $|\epsilon_{b}|$ are the wave function
and the binding energy of the isolated bound state of the CC, respectively,
and $a_{{\rm bg}}$ and $\psi_{{\rm bg}}({\bf r})$ are the scattering
length and the threshold scattering wave function of the OC in the
case without inter-channel coupling, respectivly, and 
\begin{equation}
G_{{\rm bg}}({\bf r},{\bf r}')=\frac{1}{-\left[-\nabla_{{\bf r}}^{2}+U_{oo}(r)\right]}\delta({\bf r}-{\bf r}')\label{gbg}
\end{equation}
is the Green's function for the OC in that case. It is clear that
when the result from this single-pole approximation is close to the
exact result, we can claim that the OFR is mainly due to the coupling between the OC
and the isolated bound state $\phi_b$.

In Fig.~4(a-c) we illustrate our results for three typical cases.

Case (a): 
{\it
There is no bound state in CC, and thus the OFR is induced by the scattering states of CC.
}
In Fig.~4(a) we consider the system with $a^{(+)}=1000b$ ($u^{(+)}\approx2.47b$)
and $a^{(-)}=0.5b$ ($u^{(-)}\approx-3.67b^{-2}$). Notice that here $U^{(-)}(r)$ is a repulsive square-well potential. According
to the exact numerical calculation, the OFR can occur when $\delta\approx4\times10^{-6}b^{-2}$.
Nevertheless, in this system the potentials $U_{cc}(r)=U_{oo}(r)=[U^{(+)}(r)+U^{(-)}(r)]/2$
for the OC and the CC are pure repulsive potentials. Thus, there is
no bound state in the CC. Therefore, this OFR is completely induced by
the coupling between the OC and the scattering states of the CC.  The two-body physics of Feshbach resonance induced by 
this kind of coupling
has been studied by Y. Avishai {\it et. al.}, \cite{frwobs1,frwobs2}, 
and can be understood as a kind of shape resonance of the effective
interaction in the OC. 
J. M. Acton {\it et. al.} also discussed this kind of Feshbach resonance in 
the study of many-body problem of Fermion-mediated BCS-BEC crossover \cite{ci}.
In Appendix A we provide a brief explanation
for this  kind of Feshbach resonance.

Case (b): 
{\it
There are bound states of the CC, while the OFR is induced by the scattering states of the CC.
}
In Fig.~4(b)
we consider the system with $a^{(+)}=0.15b$ ($u^{(+)}\approx19.8b^{-2}$)
and $a^{(-)}=1000b$ ($u^{(+)}\approx22.2b^{-2}$). According to the
exact numerical calculation, the OFR of this system occurs when $\delta\approx0.4\times10^{-5}b^{-2}$,
and the width of this OFR is of the order of $10^{-5}b^{-2}$. Moreover,
in this system the binding energy $|\epsilon_{b}|$ of the shallowest bound state
of the CC is as large as $14.5b^{-2}$. Therefore, when the OFR occurs,
this bound state is far off resonant to the OC. Thus, the OFR in this system is mainly due to the coupling between
the OC and the scattering states of the CC. This conclusion is further
confirmed by our calculation with the single-pole approximation, which 
shows that the scattering length $a_{{\rm s}}$ given by this approximation is significantly different from the
exact result, and does not have any resonance behavior.

Case (c): 
{\it 
The OFR is induced by the coupling between the OC and an isolated bound state of the CC.
}
In Fig.~4(c) we consider the system with $a^{(+)}=870b$ ($u^{(+)}\approx2.47b^{-2}$)
and $a^{(-)}=900b$ ($u^{(-)}\approx2.46963b^{-2}$). It is clear that for this system the results given by the single-pole approximation
and the exact numerical calculation are quantitatively consistent
with each other in the region around the OFR point. Thus, the OFR
in this system is mainly due to the coupling between the OC and the
isolated bound state of the CC.

Our results for the above cases show that the OFR with small CC-OC threshold gap
can be induced by the coupling of either type (A) or type (B) defined in Sec. I. Here
we point out that cases in Fig. 4(a) and Fig. 4(b) are essentially the same category, i.e., in the 
region of the actual OFR there is no bound state of the CC, and the OFR is induced by the coupling of type (B), i.e., the coupling between the OC and the scattering states of the CC.

We further emphasize that, as shown
in Fig. 4 (a-c), in the simple square-well model 
the OFR is usually induced by the coupling of type (A) or type (B)
when the scattering lengths
$a^{(+)}$ and $a^{(-)}$ are very close or significantly different from each other, respectively.
Nevertheless, in the ultracold
gases of alkali-earth (like) atoms, e. g., the ultracold gases of
$^{173}$Yb atoms, the realistic inter-atomic interaction potential
curve is much more complicated than the square-well model. 
For these ultracold gases, to know the OFR is induced by which type of coupling, one requires not only 
the values of $a^{(\pm)}$ but also the short-range details of the potential
curves ${U}^{(\pm)}(r)$.

\section{two-channel Huang-Yang pseudopotential}
\label{HY}

In the above section we illustrate that the OFR of the systems under the condition (\ref{con1}) can be induced 
by the coupling of two different types. In this section we show that 
in each of these two cases, the two-channel Huang-Yang pseudopotential
\cite{huangyang, ourprl} is always applicable
for the approximate calculation of the low-energy scattering amplitude.

To this end, we consider the two-atom scattering wave function $|\psi({\bf r})\rangle$ of our system, 
which can be expanded as
\begin{equation}
|\psi({\bf r})\rangle=\psi_{+}({\bf r})|+\rangle+\psi_{-}({\bf r})|-\rangle.\label{psi}
\end{equation}
This wave function satisfies the Schr$\ddot{{\rm o}}$dinger
equation 
\begin{eqnarray}
\hat{H}|\psi({\bf r})\rangle=E|\psi({\bf r}),\rangle\label{se}
\end{eqnarray} with
$E$ the scattering energy, as well as the out-going boundary condition.
As in the above discussions, we consider the systems under the condition (\ref{con1}).
We further focus on the low-energy scattering processes where the scattering energy is much smaller 
than the characteristic energy $E_{\ast}$ of the interaction potential, i.e.,
\begin{eqnarray}
E\ll E_{\ast}.\label{con2}
\end{eqnarray}
It is clear that the conditions (\ref{con1}) and (\ref{con2}) imply $1/\sqrt{E}\gg r_{\ast}$ and $1/\sqrt{\delta}\gg r_{\ast}$, where $r_{\ast}$ is the characteristic length of the interaction potential $U^{(\pm)}(r)$, as defined in Sec. II.

When $\delta=0$, the low-energy scattering wave function $|\psi({\bf r})\rangle$
has the short-range behavior 
\begin{equation}
\psi_{\pm}({\bf r})\propto\left(\frac{1}{r}-\frac{1}{a_{{\rm s}}^{(\pm)}}\right),\ r_{\ast}\lesssim r\ll\frac{1}{\sqrt{E}}\label{sb},
\end{equation}
where $a_{{\rm s}}^{(\pm)}$ is the $s$-wave scattering length with
respect to the potential $U^{(\pm)}(r)$. 

Furthermore, when $\delta$ is finite,
the behavior of the wave function in the region $r\ll1/\sqrt{\delta}$
is almost not changed. Therefore, under the conditions (\ref{con1}) and  (\ref{con2}) 
we have 
\begin{eqnarray}
\psi_{\pm}({\bf r}) & \propto & \left(\frac{1}{r}-\frac{1}{a_{{\rm s}}^{(\pm)}}\right),\ r_{\ast}\lesssim r\ll{\rm Min}\left(\frac{1}{\sqrt{E}},\frac{1}{\sqrt{\delta}}\right).\nonumber \\
\label{sb-1}
\end{eqnarray}
Due to this fact, in our calculation we can approximately replace
the real interaction potential $U^{(\pm)}(r)$ with the Bethe-Periels
boundary condition \cite{bpc} 
\begin{equation}
\lim_{r\rightarrow0}\psi_{\pm}({\bf r})\propto\left(\frac{1}{r}-\frac{1}{a_{{\rm s}}^{(\pm)}}\right),\label{bp}
\end{equation}
or the two-channel Huang-Yang pseudopotential \cite{huangyang,ourprl}
\begin{eqnarray}
\hat{U}_{{\rm HY}} & = & 4\pi\left[a_{{\rm s}}^{(+)}|+\rangle\langle+|+a_{{\rm s}}^{(-)}|-\rangle\langle-|\right]\delta({\bf r})\left(\frac{\partial}{\partial r}r\cdot\right),\nonumber \\
\label{uhy}
\end{eqnarray}
which is mathematically equivalent to the boundary condition (\ref{bp}). 
Thus, we know that for our system two-channel Huangc $\hat{U}_{{\rm HY}}$
can always be used for the approximate calculation for the low-energy two-body problems, no matter
the OFR is induced by the coupling of type (A) or (B).

To illustrate this result, we calculate the scattering length
of two atoms incident from the OC with the two-channel Huang-Yang pseudopotential $\hat{U}_{{\rm HY}}$.
To this end we consider the threshold scattering
with $E=0.$ In this case the scattering wave function given by $\hat{U}_{{\rm HY}}$
can be expressed as 
\begin{equation}
|\psi({\bf r})\rangle=\frac{1}{(2\pi)^{\frac{3}{2}}}\left[\left(1-\frac{a_{{\rm s}}^{\rm (HY)}}{r}\right)|o\rangle+B\frac{e^{-\sqrt{\delta}r}}{r}|c\rangle\right],\label{psis}
\end{equation}
where the scattering length $a_{{\rm s}}^{\rm (HY)}$
is the scattering length given by $\hat{U}_{{\rm HY}}$. The values of 
$a_{{\rm s}}^{\rm (HY)}$ 
 and the factor $B$ can
be obtained via the Schr$\ddot{{\rm o}}$dinger equation 
\begin{equation}
\left(\hat{H}_{0}+\hat{U}_{{\rm HY}}\right)|\psi({\bf r})\rangle=0.\label{se}
\end{equation}
With straightforward  calculation, we can obtain 
\begin{equation}
a_{{\rm s}}^{\rm (HY)}=\frac{-a_{{\rm s}0}+\sqrt{\delta}(a_{{\rm s}0}^{2}-a_{{\rm s}1}^{2})}{a_{{\rm s}0}\sqrt{\delta}-1},\label{as-1}
\end{equation}
where $a_{{\rm s}0}$ and $a_{{\rm s}1}$ are defined as 
\begin{equation}
a_{{\rm s}0}=\frac{1}{2}\left[a_{{\rm s}}^{(+)}+a_{{\rm s}}^{(-)}\right],\ a_{{\rm s1}}=\frac{1}{2}\left[a_{{\rm s}}^{(-)}-a_{{\rm s}}^{(+)}\right].\label{as01}
\end{equation}

In  Fig.~4(a-c) we compare $a_{\rm s}^{\rm (HY)}$ with the scattering length 
given by the exact numerical calculation. 
Notice that according to the exact numerical calculation, for these cases when
the OFR occurs the low-energy
condition (\ref{con1}) is satisfied very well ($\delta\sim10^{-6}b^{-2}-10^{-5}b^{-2}$).
As shown in Fig.~4(a-c),
in all of these cases $a_{\rm s}^{\rm (HY)}$ is very close to the result from
the exact numerical calculation, no matter if the OFR
is induced by the coupling between the bound or scattering states
of the CC. 

\begin{figure}
\includegraphics[width=9cm]{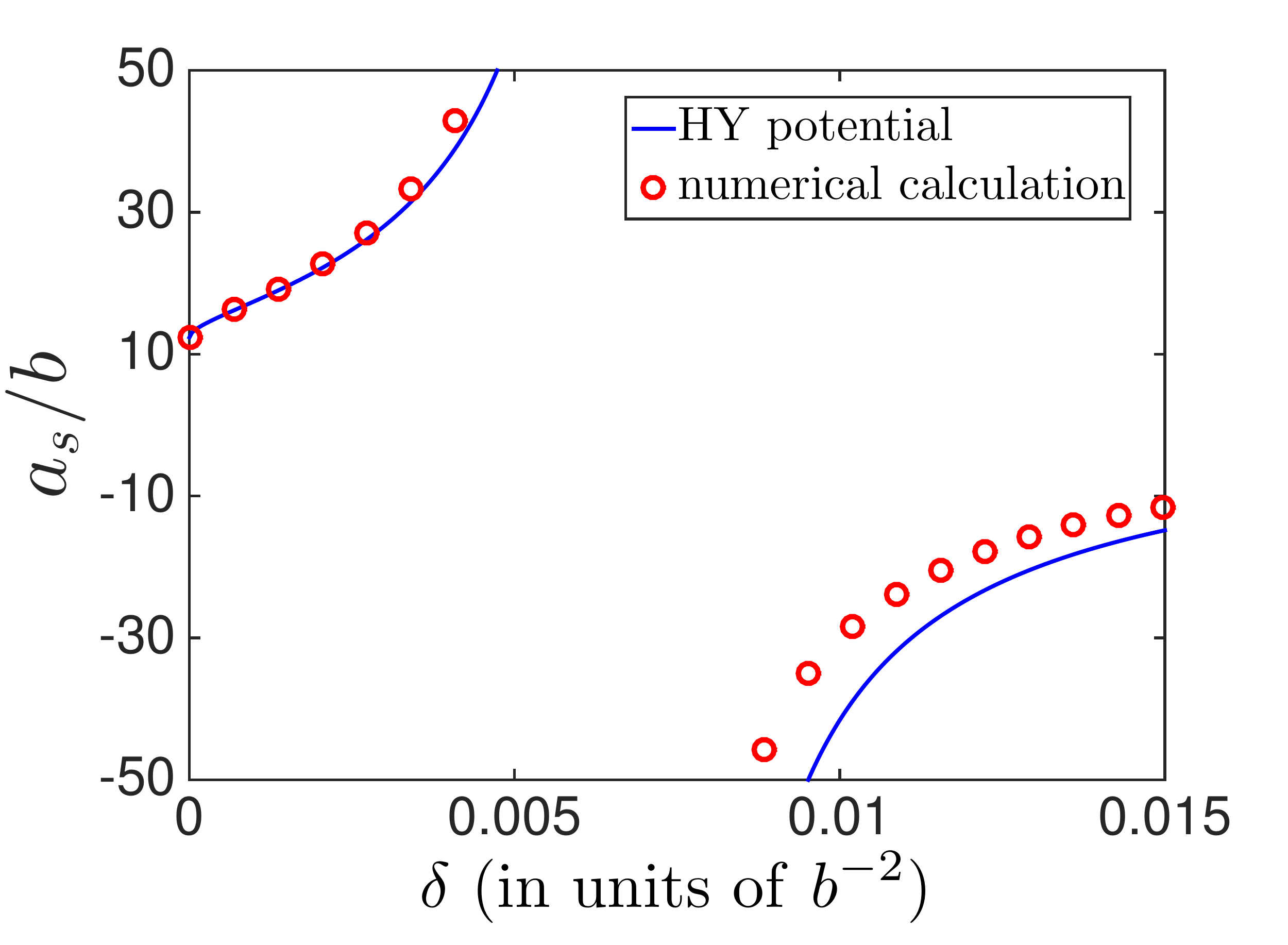} 
 \caption{(color online) The scattering length $a_{{\rm s}}$ of square-well
model. We show the results given by exact numerical calculation
(red circle) and the two-channel Huang-Yang pseudopotential (blue
solid line). Here we consider the cases with $a^{(+)}=22.4b$
($u^{(+)}\approx2.56b^{-2}$) and $a^{(-)}=2.34b$ ($u^{(-)}\approx3.76b^{-2}$).}
\end{figure}

We also do a claculation for the $^{{\rm 173}}$Yb atoms with our
square well model and the two-channel Huang-Yang pseudo potential.
In our calculation we take $b=R_{{\rm vdW}}=84.84a_{0}$ \cite{spinexchange3},
$a^{(+)}=1900a_{0}\approx22.4b$ ($u^{(+)}\approx2.56b^{-2}$) and
$a^{(-)}=200a_{0}\approx2.34b$ ($u^{(-)}\approx3.76b^{-2}$) \cite{ofrexp1,ofrexp2}.
We illustrate our results in Fig. 5. It is shown that although
we can still approximately derive the scattering length $a_{{\rm s}}$
with the Huang-Yang pseudopotential $\hat{U}_{{\rm HY}}$, there is
a relative error of $10\%$ in this approximation. This fact can be
understood with the following analysis. For $^{{\rm 173}}$Yb atoms,
the OFR occurs when $\delta\sim10^{-2}b^{-2}$. Thus, in this system
the low-energy condition (\ref{con1}) is not satisfied as perfectly
as in the examples of Fig. 4(a-c). As a result, the relative error
of the pseudopotential approximation is larger. Our result implies
that to obtain more accurate theoretical result for $^{{\rm 173}}$Yb
atoms , one needs to include more details of the interaction $U^{(\pm)}(r)$
in our calculation.

\section{Discussion}
\label{sec:dis}
In this paper we show that the OFR in the systems where the CC-OC threshold
gap $\delta$ is much smaller than the characteristic energy $E_{\ast}$ 
can be induced
by the coupling between the OC and either an isolated bound state
or the scattering states of the CC. In any case, the two-channel Huang-Yang pseudopotential $\hat{U}_{{\rm HY}}$ can always be used as a good approximation for
the inter-atomic interaction potential in the region $\delta\ll E_{\ast}$. We illustrate these conclusions
with a simple square-well model.


According to Eq. (\ref{as-1}), the scattering length $a_{{\rm s}}^{\rm (HY)}$
given by the Huang-Yang pseudo potential diverges when 
\begin{eqnarray}
\delta=\frac{1}{a_{\rm s0}^2},\label{ofrc}
\end{eqnarray}
with $a_{\rm s0}$ defined in Eq. (\ref{as01}).
Therefore, the OFR can occur in the systems with 
$\delta\ll E_{\ast}$ only when $1/a_{\rm s0}^2\ll E_{\ast}$, i.e.,
$a_{\rm s0}$ should be much larger than the characteristic length $r_{\ast}$ of the interaction potential.
According to Eq. (\ref{east}) and Eq. (\ref{as01}), this condition yields that either of the two 
scattering lengths $a_{\rm s}^{(+)}$ and $a_{\rm s}^{(-)}$
for the channels $|+\rangle$ and $|-\rangle$
 should be positive and much larger than $r_{\ast}$. Therefore, 
 when 
an OFR occurs
under the condition $\delta\ll E_{\ast}$, there must be a
shallow bound state in the channel $|+\rangle$ or $|-\rangle$,
although there may be no
shallow 
 bound state in the CC $|c\rangle$.

Here we also would like to make a comment for the scattering length of the CC itself.
Naively, by projecting  the two-channel  Huang-Yang pseduopotential
$\hat{U}_{{\rm HY}}$ on the CC, one can obtain
\begin{equation}
\langle c|\hat{U}_{{\rm HY}}|c\rangle=4\pi a_{{\rm s}0}\delta({\bf r})\left(\frac{\partial}{\partial r}r\cdot\right).\label{uhyc}
\end{equation}
Thus, it seems that the scattering length of the CC is just $a_{{\rm s}0}=(a_{{\rm s}}^{(+)}+a_{{\rm s}}^{(-)})/2$.
However, this conclusion is incorrect. 
The real
interaction potential of the CC is 
\begin{equation}
\langle c|U|c\rangle=U_{cc}(r)=\frac{1}{2}\left[U^{(+)}(r)+U^{(-)}(r)\right].\label{uc}
\end{equation}
It is clear that the scattering length of this real potential, which
can be denoted as $a_{{\rm sc}}$, is determined by not only the scattering
length $a_{{\rm s}}^{(\pm)}$ of the potential curves $U^{(\pm)}(r)$,
but also the short-range details $U^{(\pm)}(r)$. The value of $a_{{\rm sc}}$
is possible to be either significantly different from $a_{{\rm s}0}$
or close to $a_{{\rm s0}}$. Accordingly, although the complete two-channel
Huang-Yang pseudopotential $\hat{U}_{{\rm HY}}$ is always applicable
in a\emph{ }two-channel problem\emph{ }under the conditions
(\ref{con1}) and (\ref{con2}), it does not means that we can directly use the projection
$\langle c|\hat{U}_{{\rm HY}}|c\rangle$ to study
the\emph{ }single-channel\emph{ }physics for the CC itself.

As shown above, the OFR can
occur in the region $\delta\ll E_{\ast}$ in a system with $a_{\rm s0}>0$ and $1/a_{\rm s0}^2\ll E_{\ast}$.
In such a system, if the value of the scattering length $a_{\rm sc}$ of the CC 
is close to the value of $a_{\rm s0}$, then there exists a shallow bound state in the CC with
binding energy 
close to $1/a_{\rm s0}^2$. When the OFR occurs, i.e. $\delta\approx1/a_{\rm s0}^2$,
this bound state is near resonant to the threshold of the OC, and 
thus has a significant contribution to the OFR. On the other hand, if the value of 
$a_{\rm sc}$ is negative or much smaller than $a_{\rm s0}$, then when the OFR
occurs all the bound states of the CC are far off resonant to the threshold of the OC.
In that system the OFR may be mainly induced by the coupling between the OC and the 
scattering states of the CC.

 It is also pointed out that, in principle
 the Feshbach resonance induced by the coupling 
 between the OC and the scattering states of the CC can also occur
in the systems where the CC-OC threshold
gap is comparable or larger than the characteristic energy \cite{frwobs1,frwobs2}. 
However, in these systems that kind of Feshbach resonance
usually requires extremely strong CC-OC coupling 
\cite{frwobs1,frwobs2}, which is very difficult to be
generated in the realistic systems. Therefore, 
the Feshbach resonances in realistic ultrcold gases with large CC-OC threshold
gap, e.g., the MFR of ultrcold alkali atoms, are usually induced by the 
coupling 
 between the OC and one or several isolated bound states of the CC.

\begin{acknowledgments}
We thank Yshai Avishai, Jesper Levinsen and Hui Zhai for helpful discussions. We also thank Hui Zhai for reading the manuscript.
This work has been supported by National Natural Science Foundation
of China under Grants No. 11222430 and No. 11434011, NKBRSF of China
under Grants No. 2012CB922104. This work is also supported by the Fundamental Research Funds for the
Central Universities, and the Research Funds of Renmin University of China 16XNLQ03.


\end{acknowledgments}
\addcontentsline{toc}{section}{Appendices}\markboth{APPENDICES}{}

\appendix

\section{Feshbach Resonance Induced by the Scattering States of the CC}

In this appendix we show that the Feshbach resonance induced by the
coupling between the OC and the scattering states of the CC can be
understood as a shape resonance of the effective interaction in the
OC. For convenience, here we also consider the OFR system discussed
in our main text. Nevertheless, our result is applicable for the general 
two-channel scattering problems of two atoms.

As shown in Sec. II, the Hamiltonian of our system can be expressed as
\begin{eqnarray}
\hat{H} & = & \left(-\nabla_{{\bf r}}^{2}+\delta\right)|c\rangle\langle c|+\left(-\nabla_{{\bf r}}^{2}\right)|o\rangle\langle o|+\hat{U}\nonumber \\
 & \equiv & \hat{H}_{0}+\hat{U},\label{ah}
\end{eqnarray}
where the interaction potential
$\hat{U}$ can be written as $\hat{U}=\sum_{i,j=o,c}U_{ij}(r)|i\rangle\langle j|$,
with $U_{oo}(r)$ and $U_{cc}(r)$ being the intra-channel interaction
potential of the OC and CC, respectively, and $U_{oc}(r)=U_{co}(r)^{\ast}$
being the inter-channel coupling.

In our system the scattering wave function can be expressed as 
\begin{equation}
|\psi({\bf r})\rangle=\psi^{(o)}({\bf r})|o\rangle+\psi^{(c)}({\bf r})|c\rangle\label{apsi}
\end{equation}
and satisfies the Schr$\ddot{{\rm o}}$dinger equation 
\begin{equation}
\hat{H}|\psi({\bf r})\rangle=E|\psi({\bf r})\rangle,\label{ase1}
\end{equation}
with $E$ the scattering energy. Furthermore, when $E=0$ we have
\begin{equation}
\lim_{r\rightarrow\infty}\psi^{(o)}({\bf r})\propto\left(1-\frac{a_{\rm s}}{r}\right),\label{abp}
\end{equation}
where $a_{\rm s}$ is the scattering length of two atoms incident from the
OC. Therefore, to obtain the scattering length $a_{\rm s}$, we should first
calculate the component $\psi^{(o)}({\bf r})$ of the scattering wave
function for the case with $E=0$. In this case, Eq. (\ref{ase1})
can be re-written as 
\begin{eqnarray}
 &  & -\nabla_{{\bf r}}^{2}\psi^{(c)}({\bf r})+\delta\psi^{(c)}({\bf r})+U_{cc}(r)\psi^{(c)}({\bf r})\nonumber \\
 &  & +U_{co}(r)\psi^{(o)}({\bf r})=0;\label{aec}\\
 &  & -\nabla_{{\bf r}}^{2}\psi^{(o)}({\bf r})+U_{oo}(r)\psi^{(o)}({\bf r})+U_{oc}(r)\psi^{(c)}({\bf r})=0.\nonumber \\
\label{aeo}
\end{eqnarray}

Using these two equations, we can obtain the effective interaction
potential of the OC via the projection operator technique. To this
end we first solve Eq. (\ref{aec}) and obtain 
\begin{equation}
\psi^{(c)}({\bf r})=\int d{\bf r}'g(\delta,{\bf r},{\bf r}')U_{co}(r')\psi^{(o)}({\bf r}'),\label{apsic}
\end{equation}
where $g(\delta,{\bf r},{\bf r}')$ is the Green's function of the
CC, and is defined as 
\begin{equation}
g(\delta,{\bf r},{\bf r}')=\frac{1}{-\left[-\nabla_{{\bf r}}^{2}+U_{cc}(r)+\delta\right]}\delta({\bf r}-{\bf r}').\label{eq:}
\end{equation}
We assume all the bound states of the CC is far off resonant to the
threshold of the OC. Thus, in the calculation of $g(\delta,{\bf r},{\bf r}')$
we can neglect the contributions from the bound states of the CC.
Then we obtain 
\begin{equation}
g(\delta,{\bf r},{\bf r}')\approx\int d{\bf k}\frac{\psi_{{\bf k}}^{\ast}({\bf r})\psi_{{\bf k}}({\bf r}')}{-\delta-k^{2}},\label{ag2}
\end{equation}
where $\psi_{{\bf k}}({\bf r})$ is the scattering state of the CC
with incident momentum ${\bf k}$. Substituting Eq. (\ref{apsic})
into Eq. (\ref{aeo}), we obtain the equation for $\psi^{(o)}({\bf r})$:
\begin{equation}
-\nabla_{{\bf r}}^{2}\psi^{(o)}({\bf r})+{\cal V}_{{\rm eff}}(\delta)\psi^{(o)}({\bf r})=0.\label{ase-1}
\end{equation}
Here ${\cal V}_{{\rm eff}}(\delta)$ is the effective interaction
of the OC. It is non-diagonal in the ${\bf r}$-representation, and
satisfies 
\begin{eqnarray}
 &  & {\cal V}_{{\rm eff}}(\delta)\psi^{(o)}({\bf r})\nonumber \\
 & = & U_{oc}(r)\int d{\bf r}'g(\delta,{\bf r},{\bf r}')U_{co}(r')\psi^{(o)}({\bf r}')\nonumber \\
 &  & +U_{oo}(r)\psi^{(o)}({\bf r})\label{veff}
\end{eqnarray}
with $g(\delta,{\bf r},{\bf r}')$ given by Eq. (\ref{ag2}).

Eq. (\ref{ase-1}) and Eq. (\ref{abp}) show that we can obtain the
wave function $\psi^{(o)}({\bf r})$ and the scattering length $a_{\rm s}$
of our two-channel scattering problem via solving the single-channel
scattering problem with effective interaction ${\cal V}_{{\rm eff}}(\delta)$,
which changes with the CC-OC threshold gap $\delta$. Although
in ${\cal V}_{{\rm eff}}(\delta)$ we have ignored all the contributions
from the bound states of the CC, it is still possible that a shape
resonance for ${\cal V}_{{\rm eff}}(\delta)$ can appear when $\delta$
is tuned to a particular value $\delta_{0}$ ($\delta_{0}>0$). In
this case, we would have $a_{\rm s}=\infty$. It is clear that this resonance
is nothing but the Feshbach resonance induced by the coupling between
the OC and the scattering states of the CC.


\begin{thebibliography}{10}
\bibitem{fr} H. Feshbach, Ann. Phys. {\bf 5}, 357 (1958).

\bibitem{chinrmp} C. Chin, R. Grimm, P. Julienne and E. Tiesinga, Rev. Mod. Phys. {\bf 82}, 1225 (2010).

\bibitem{paulrmp} T. K\"{o}hler, K. G\'{o}ral and P. S. Julienne, Rev. Mod. Phys. {\bf 78}, 1311 (2006).

\bibitem{ourprl} R. Zhang, Y. Cheng, H. Zhai, and P. Zhang, Phys. Rev. Lett. {\bf 115}, 135301 (2015).

\bibitem{spinexchange1} G. Cappellini, M. Mancini, G. Pagano, P. Lombardi, L. Livi, M. Siciliani de Cumis, P. Cancio, M. Pizzocaro, D. Calonico, F. Levi, C. Sias, J. Catani, M. Inguscio, and L. Fallani, Phys. Rev. Lett. {\bf 113}, 120402 (2014) and Phys. Rev. Lett. {\bf 114}, 239903 (2015).

\bibitem{spinexchange2}F. Scazza, C. Hofrichter, M. H\"ofer, P. C. De Groot, I. Bloch, and S. F\"olling, Nature Phys. {\bf 10}, 779 (2014) and Nature Phys. {\bf 11}, 514 (2015).

\bibitem{spinexchange3} X. Zhang, M. Bishof, S. L. Bromley, C. V. Kraus, M. S. Safronova, P. Zoller, A. M. Rey, J. Ye., Science, {\bf 345}, 1467 (2014). 

\bibitem{zeeman} M. M. Boyd, T. Zelevinsky, A. D. Ludlow, S. Blatt, T. Willette, S. M. Foreman, and J. Ye, Phys. Rev. A \textbf{76}, 022510 (2007). 

\bibitem{ofrexp1} G. Pagano, M. Mancini, G. Cappellini, L. Livi, C. Sias, J. Catani, M. Inguscio, L. Fallani,  Phys. Rev. Lett. {\bf 115}, 265301 (2015).

\bibitem{ofrexp2} M. H\"ofer, L. Riegger, F. Scazza, C. Hofrichter, D.R. Fernandes, M. M. Parish, J. Levinsen, I. Bloch, S. F\"olling,  Phys. Rev. Lett. {\bf 115}, 265302 (2015).

\bibitem{huangyang} K. Huang and C. N. Yang, Phys. Rev. {\bf 105}, 767 (1957).

\bibitem{frwobs1} Y. Avishai, Y. B. Band, and M. Trippenbach, Phys. Rev. Lett. {\bf 111}, 155301 (2013).

\bibitem{frwobs2} T. Wasak, M. Krych, Z. Idziaszek, M. Trippenbach, Y. Avishai, and Y. B. Band, Phys. Rev. A {\bf 90}, 052719 (2014).






\bibitem{bpc} H. Bethe and R. Peierls, Proc. R. Soc. A {\bf 148}, 146 (1935).


\bibitem{swpaper1} S. J. J. M. F. Kokkelmans, J. N. Milstein, M. L. Chiofalo, R. Walser, and M. J. Holland, Phys. Rev. A {\bf 65},
053617 (2002).

\bibitem{swpaper2} C. Chin, arXiv:cond-mat/0506313.
\bibitem{swpaper3} A. D. Lange, K. Pilch, A. Prantner, F. Ferlaino, B. Engeser, H.-C. N${\ddot {\rm a}}$gerl, R. Grimm, and C. Chin, Phys. Rev. A {\bf 79}, 013622 (2009).

\bibitem{ci} J. M. Acton, M. M. Parish, and B. D. Simons, Phys. Rev. A {\bf 71},
063606 (2005).


\end{thebibliography}
\end{document}